\documentclass[12pt,thmsa]{article}
\usepackage{sw20nhj}

\input{tcilatex}
\begin{document}

\author{\textbf{V. Akulov and M. Kudinov} \\
\textit{Physics Department,} \textit{Baruch College of the City University
of New York}\\
\textit{New York, NY10010, USA }\\
\textit{e-mail: akulov@gursey.baruch.cuny.edu}\\
\textit{e-mail: kudinov@gursey.baruch.cuny.edu}}
\title{Extended Supersymmetric Quantum Mechanics}
\date{May 10, 1999 }
\maketitle

\begin{abstract}
A parametrization of the Hamiltonian of the generalized Witten model of the
SUSY QM by a single arbitrary function in $d=1$ has been obtained for an
arbitrary number of the supersymmetries $\mathcal{N}$\textrm{\ }$=2N$.
Possible applications of this formalism have been discussed. It has been
shown that the $N=1$ and $2$ conformal SUSY QM is generalized for any $N$.

\hspace{21.4in}


\hspace{15.4in}\hspace{15.4in}\hspace{0.66in}\hspace{1.75in}\hspace{5.4in}
\end{abstract}

\section{Introduction}

Recently, the interest to the supersymmetric quantum mechanics \cite{Witten}%
, \cite{Ritenberg} has been renewed. Following the Maldacena's conjecture 
\cite{Maldacena} it has been argued \cite{Kalosh1}, that the extended
superconformal quantum mechanics \cite{Akulov}, \cite{Fubini}, \cite{Ivanov}%
, \cite{Azscaraga} describes the dynamics of a $D-0$ brane in the near
horizon background of the extreme Reisner-Nordstrom black hole. In the
framework of the $N=1$ SUSY QM the exactly solvable potentials of the common
QM were classified and found to be related to the integrability condition
called shape invariance \cite{Gendenshtein}, \cite{PysRep95}. There is an
exciting connection of the extended SUSY QM to the solution of the Calogero
type many body potentials \cite{Freedman}. It has been noted \cite{Gibbons},
that the $\mathcal{N}\geq 4$ superconformal Calogero model has not been
constructed yet. These motivations lead us to consider an old problem of
extending the SUSY QM to an arbitrary number of the supersymmetries $N$.
Unlike the supersymmetric field theories, where the number $\mathcal{N}$ is
constrained by the spin content ( $\mathcal{N}=4$ for the SYM, and $\mathcal{%
N}=8$ for the SUGRA both in $D=4$), the extended SUSY QM appears to have
none. Thus, we can consider $\mathcal{N}$as a free parameter in the theory.
For the same reason the extended SUSY QM of $\mathcal{N}>8$ can not be the
result of the dimensional reduction of the corresponding field theory.
Originally, the SUSY QM was first considered in the algebraic approach by
Witten \cite{Witten}, \cite{Ritenberg} before the superfield approach was
developed in \cite{Akulov}, \cite{CoopFree}. The main advantage of the
latter is that it gives the SUSY QM a natural classical foundation. On the
other hand, the algebraic approach is more directly related to the quantum
mechanical aspects of the theory. Although, it is possible to obtain the
results of this paper in both formalisms, we will consider here only the
generalization of the Witten model, leaving the quasiclassical superfield
treatment for elsewhere. The Hamiltonian of the SUSY QM has been obtained by
Pashnev \cite{Pashnev} for $N=2$ and $3$. It was parametrized by a single
arbitrary function known as the superpotential. The case of an arbitrary $N$
has not been considered there or anywhere else. In a similar approach we
have found a framework for extending the SUSY QM to an arbitrary $N$. Our
main assumption is that the Hamiltonian of the SUSY QM\ must be the sum of
the irreducible representations of the symmetric group $S_N$, which
transform the supercharges. The representations can be parametrized by some
functions, which are all related through a compatibility condition to a
single arbitrary parameter. The compatibility condition is the result of the
requirement for the representations to be totally symmetric and to satisfy
superalgebra. It consists of the system of $N-1$ first order non-linear
differential equation for $N$ functions, which can be solved in general for
any $N$. In the next sections we elaborate on the general formalism and as
an illustration consider $N=2$ and $3$ cases. Then we treat the arbitrary $N$
case. In the end as a special example, we consider the extension of the
conformal SUSY\ QM to an arbitrary $N$.

\section{General formalism}

The $N$ SUSY QM is defined by the generators $Q_i,$ $Q_i^{+},$ $H$ of the
superalgebra in the simplest case without the central charges as follows 
\begin{eqnarray}
\{Q_i,Q_j^{+}\} &=&H\delta
_{ij},\{Q_i,Q_j\}=\{Q_i^{+},Q_j^{+}\}=0,[Q_i,H]=[Q_i^{+},H]=0  \label{N} \\
Q_i &=&\sigma _i^{-}\left( \hbar \partial _x+W(x,\sigma _1,...,\sigma
_{i-1},\sigma _{i+1},...,\sigma _N)\right)  \nonumber \\
H &=&-\hbar ^2\partial _x^2+W^2+\hbar \sigma _iW^{\prime }  \label{V}
\end{eqnarray}
where $\sigma _i^{+},\sigma _i^{-\ },\sigma _i^3$, $i=1...N$ are sets of the
Pauli matrices with $[\sigma _i^{+},\sigma _i^{-}]=$ $\sigma _i^3$. In
addition, they satisfy the Clifford algebra $\{\sigma _i^{+},\sigma
_j^{-}\}=\delta _{ij}$ and represent the fermionic degrees of freedom,
appearing naturally in the superfield approach after the quantization.%
\footnote{%
The Clifford algebra $C_N$ can be realized in a standard way in terms of the
direct product of the $\sigma $ and unit matrices: 
\[
\sigma _i^3\equiv \sigma _i=1_1\otimes ...\otimes 1_{i-1}\otimes \sigma
^3\otimes 1_{i+1}\otimes ...\otimes 1_N 
\]
} We require that the $N$ supercharges are to be transformed by the
irreducible representations of the symmetric group. Hence, a superpotential $%
W$ and its derivative $W^{\prime }$ must be totally symmetric with respect
to its fermionic variables. The form of the expression for $N$ supercharges
above insures that the anticommutators of $Q_i$ and $Q_j^{+}$ vanish for all 
$i\neq j$. Another requirement of the superalgebra (\ref{N}), that is the
anticommutators of $Q_i$ and $Q_i^{+}$give the same Hamiltonian for all $i$
is sattisfied by imposing the equation 
\begin{eqnarray}
&&W^2(x,\sigma _2,\sigma _3,...,\sigma _N)+\hbar \sigma _1W^{\prime
}(x,\sigma _2,\sigma _3,...,\sigma _N)  \label{compatcon} \\
&=&W^2(x,\sigma _1,\sigma _3,...,\sigma _N)+\hbar \sigma _2W^{\prime
}(x,\sigma _1,\sigma _3,...,\sigma _N)  \nonumber
\end{eqnarray}

\smallskip

on the potentials in (\ref{V}). In fact, since the $W$ is totally symmetric
in all $\sigma _i$, the equation for the two potentials written above is
sufficient.We call this a compatibility condition. The explicitly symmetric
form of the $W$ is written as 
\begin{equation}
W(x,\sigma _2,\sigma _3,...,\sigma _N)=W_0+(\sigma _2+\sigma _3+...+\sigma
_N)W_1+...+\sigma _2\sigma _3...\sigma _NW_{N-1}  \label{WN}
\end{equation}
With this $W$ the compatibility condition (\ref{compatcon}) turns into a
system of $N-1$ first order differential equations for $N$ functions $W_0$,$%
W_1$,...,$W_{N-1}$. Thus, the solution and the Hamiltonian can be
parametrized by an arbitrary function.

\section{N=2 and 3}

It is now appropriate to illustrate the general formalism in simple cases,
as well as to establish the connection with the known results.

\subsection{N=2}

The superpotential $W$ and the compatibility condition are 
\begin{eqnarray}
W_{N=2}(x,\sigma _2) &=&W_0+\sigma _2W_1  \label{Wformn=2} \\
W^2(x,\sigma _2)+\hbar \sigma _1W^{\prime }(x,\sigma _2) &=&W^2(x,\sigma
_1)+\hbar \sigma _2W^{\prime }(x,\sigma _1)  \nonumber
\end{eqnarray}
The equation on $W_0$ and $W_1$ is 
\begin{equation}
\hbar W_0^{\prime }=2W_0W_1  \label{Eqn=2}
\end{equation}
The Hamiltonian $H$ and the superpotential are parametrized by $W_0$ as 
\begin{eqnarray}
H_{N=2} &=&-\hbar ^2\partial _x^2+\left( W_0+\sigma _2\frac \hbar 2\frac{%
W_0^{\prime }}{W_0}\right) ^2+\hbar \sigma _1\left( W_0+\sigma _2\frac \hbar
2\frac{W_0^{\prime }}{W_0}\right) ^{\prime }  \label{Hn=3} \\
W_{N=2}(x,\sigma _2) &=&W_0+\sigma _2\frac \hbar 2\frac{W_0^{\prime }}{W_0}
\end{eqnarray}
This reduces to $N=1$ case by setting $\sigma _2$ to zero.

\subsection{N=3.}

The superpotential $W$ can be parametrized recursively as 
\begin{equation}
W_{N=3}(x,\sigma _2,\sigma _3)=W_0+\sigma _2W_1+\sigma _3\left( W_1+\sigma
_2W_2\right) .  \label{WN=3}
\end{equation}
Comparing this with $W_{N=2}$ (\ref{Wformn=2}), suggests that the
compatibility condition can be written as 
\begin{equation}
\hbar \left( W_0(x)+\sigma _2W_1(x)\right) ^{\prime }=2\left( W_0(x)+\sigma
_2W_1(x)\right) \left( W_1(x)+\sigma _2W_2(x)\right)  \label{Eqn=3}
\end{equation}
Introducing new variables $w_1=W_0+W_1$ and $w_2=W_0-W_1$ with two other
relations $\frac \hbar 2\frac{w_1^{\prime }}{w_1}=$ $W_1+W_2$ and $\frac
\hbar 2\frac{w_2^{\prime }}{w_2}=$ $W_1-W_2$ it reduces to the equation%
\footnote{%
The case when $w_1$or $w_2$ equals zero must be treateed separately.} 
\begin{equation}
w_1-w_2=\frac \hbar 2\frac{w_1^{\prime }}{w_1}+\frac \hbar 2\frac{%
w_2^{\prime }}{w_2}\text{.}  \label{w1w2eq}
\end{equation}
With $w_1=$ $w_2+\frac \hbar 2\frac{u^{\prime }}u$ in (\ref{w1w2eq}), $u$
being an arbitrary function, it turns into the quadratic equation 
\begin{equation}
w_1^2-\frac \hbar 2\frac{u^{\prime }}uw_1-u=0  \label{quadEQ}
\end{equation}
with two roots for $w_1$ and $w_2$%
\begin{equation}
w_1=\frac \hbar 4\frac{u^{\prime }}u\pm \frac \hbar 4\sqrt{\left( (\frac{%
u^{\prime }}u)^2+16u\right) \text{ }}\text{\quad }w_2=-\frac \hbar 4\frac{%
u^{\prime }}u\pm \frac \hbar 4\sqrt{\left( (\frac{u^{\prime }}%
u)^2+16u\right) \text{ }}  \label{R}
\end{equation}
The superpotential $W_{N=3}$ is parametrized through its function
coefficients in terms of $u$ as 
\begin{equation}
W_0=\frac \hbar 4\sqrt{\left( (\frac{u^{\prime }}u)^2+16u\right) \text{ }}%
\text{ \quad }W_1=\frac \hbar 4\frac{u^{\prime }}u\text{\quad }W_2=\frac
\hbar 4\frac{2u^{\prime \prime }u-3(u^{\prime })^2}{u^2\sqrt{\left( (\frac{%
u^{\prime }}u)^2+16u\right) }}  \label{Ws}
\end{equation}

This is the superpotential that was found by Pashnev \cite{Pashnev}. There
is a different parametrization of the Hamiltonian, which is more useful in
dealing with the extension to an arbitrary $N$. The equation (\ref{w1w2eq})
can be written as 
\begin{equation}
\hbar w_2^{\prime }=-2w_2^2+w_2{\LARGE (}2w_1-\hbar \frac{w_1^{\prime }}{w_1}%
{\LARGE )}  \label{Ricati}
\end{equation}
This is a Bernoulli's equation with the general solution 
\begin{equation}
w_2=\frac \hbar 2\ln {}^{\prime }{\Huge (}\int \frac{e^{\frac 2\hbar \int
w_1}}{w_1}{\Huge )}  \label{RicSol}
\end{equation}
and $w_1$ taken as a parameter. This case can be also reduced to $N=2$ by
setting $\sigma _3$ equal to zero.

\section{Arbitrary N}

The symmetric superpotential $W(x,\sigma _2,\sigma _3,...,\sigma _N)$ is
parametrized recursively by $N$ functions $W_0$,$W_1$, ..., $W_{N-1}$%
\begin{eqnarray}
W &=&W_0+\sigma _2W_1+\sigma _3\left( W_1+\sigma _2W_2\right) +...+\sigma _N%
{\LARGE (}W_1+\sigma _2W_2+\sigma _3\left( W_2+\sigma _2W_3\right)  \nonumber
\\[0.12in]
&&+...+\sigma _{N-1}{\Large (}...+\sigma _3\left( W_{N-2}+\sigma
_2W_{N-1}\right) {\Large )}...{\LARGE )}  \label{aa}
\end{eqnarray}
The new variables $w_1$, $w_2$, ... ,$w_{N-1}$ are introduced similarly to $%
N=$ $3$ case. They are all the distinct combinations of the functions $W_0$,$%
W_1$, ..., $W_{N-2}$ in the first half of the superpotential $W$ as $\sigma
_i$ run over the values $\pm 1$ in turn. We have other $N-1$ relations
between $W_1$,$W_2$, ..., $W_{N-1}$ from the compatibility condition in the
way similar to (\ref{Eqn=3}). In the new variables $w_i$ the compatibility
condition is written as the system of $N-2$ equations 
\begin{eqnarray}
w_1-w_2 &=&\frac \hbar 2\frac{w_1^{\prime }}{w_1}+\frac \hbar 2\frac{%
w_2^{\prime }}{w_2}\text{ \qquad }w_2-w_3=\frac \hbar 2\frac{w_2^{\prime }}{%
w_2}+\frac \hbar 2\frac{w_3^{\prime }}{w_3}  \nonumber \\
\text{... }w_{N-2}-w_{N-1} &=&\frac \hbar 2\frac{w_{N-2}^{\prime }}{w_{N-2}}%
+\frac \hbar 2\frac{w_{N-1}^{\prime }}{w_{N-1}}
\end{eqnarray}
The solution of a given equation is expressed in terms of the solution of
the previous one according to the iteration scheme

\begin{equation}
w_2=\frac \hbar 2\ln ^{\prime }\left( \int \frac{e^{\frac 2\hbar \int w_1}}{%
w_1}\right) \text{ }w_3=\frac \hbar 2\ln ^{\prime }\left( \int \frac{%
e^{\frac 2\hbar \int w_2}}{w_2}\right) \text{... }w_{N-1}=\frac \hbar 2\ln
^{\prime }\left( \int \frac{e^{\frac 2\hbar \int w_{N-2}}}{w_{N-2}}\right)
\end{equation}
For any $N$ it is always possible to invert a system of linear equations in
order to express $W_i$ in terms of $w_i$, so that the Hamiltonian is
parametrized by a single function, say $w_1$.

\section{Superconformal QM and Calogero Type Potentials}

The conformal quantum mechanics in $d=1$ is defined \cite{Alfaro} by the
generators of the conformal group $H$, $K$, and $D$. The Hamiltonian of the
theory 
\begin{equation}
~H=\frac 12\left( p^2+\frac g{x^2}\right)  \label{Conformal}
\end{equation}
was found to have a non-normalizable ground state. When the isomorphism
between the conformal group and the group $SO(2,1)$ was used to construct
new generators as the linear combinations of the old, it was suggested to
use the generator of the compact rotation of $SO(2,1)$ as a new Hamiltonian
( with $\hbar =1$ ) 
\begin{equation}
H_{new}=\frac 12\left( -\partial ^2+\frac g{x^2}+\frac{x^2}a\right)
\label{Hnew}
\end{equation}
which has a normalizable ground state and the spectrum similar to that of a
compact operator 
\begin{eqnarray}
r &=&r_0+n\text{, }n=0,1,...  \label{Spectr} \\
r_0 &=&\frac 12(1+\sqrt{g+\frac 14})  \nonumber
\end{eqnarray}
It has been noted, that this is equivalent to elimination of a coordinate
singularity \cite{Kalosh2} or to a non-linear change of the space-time
variables \cite{Akulov}. In the later developments \cite{Akulov}, \cite
{Fubini} the supersymmetric conformal QM for $N=1$ and $2$ was constructed.
After regularization similar to that in (\ref{Hnew}) the spectrum of the $%
N=1 $ Hamiltonian 
\begin{equation}
H_{N=1}=\frac 12\left( -\partial _x^2+\frac{\lambda ^2}{x^2}-\sigma _1\frac
\lambda {x^2}\right)  \label{N1Ham}
\end{equation}
can be obtained from the that of the common conformal QM (\ref{Spectr}) by
the substitution $g\rightarrow \lambda (\lambda -s1)$, with $s1=\pm 1$ being
the eigenvalues of the $\sigma _1$%
\begin{eqnarray}
r &=&\frac 12(\lambda +\frac 12)+n\text{\qquad \qquad }s1=+1
\label{SpectrN=1} \\
r &=&\frac 12(\lambda +\frac 12)+\frac 12+n\text{\qquad }s1=-1  \nonumber
\end{eqnarray}

Notice, that the $x$-dependence is factorized in the conformal potential,
leaving the numerator as a polynomial of the fermionic variable $\sigma $.
It has been conjectured \cite{Pashnev}, that the factorization of the
bosonic variables in the case of the conformal potential persists for an
arbitrary $N$. We are going to show next that this is indeed the case.
Moreover, we have obtained a whole class of the potentials with such
property. The number of the potentials in each class grows with $N$. In
addition to the requirement for the Hamiltonian to be a totally symmetric
form of its fermionic variables we also demand the factorization of its
bosonic variables. That leads us to the following ansatz for the
superpotential $W$

\begin{eqnarray}
W(x,\sigma _2,\sigma _3,...,\sigma _N) &=&W_0P(\sigma _2,\sigma
_3,...,\sigma _N)  \label{anz} \\
P(\sigma _2,\sigma _3,...,\sigma _N) &=&1+a_1(\sigma _2+...+\sigma
_N)+...+a_{N-1}\sigma _2\sigma _3...\sigma _N
\end{eqnarray}
where all the $W_i$ are proportional to $W_0$ with coefficients $a_i$. The
compatibility condition (\ref{compatcon}) is now 
\begin{eqnarray}
&&W_0^2P(\sigma _2,\sigma _3,...,\sigma _N)^2+\sigma _1W_0^{\prime }P(\sigma
_2,\sigma _3,...,\sigma _N)  \label{compat1} \\
&=&W_0^2P(\sigma _1,\sigma _3,...,\sigma _N)^2+\sigma _2W_0^{\prime
}P(\sigma _1,\sigma _3,...,\sigma _N)  \nonumber
\end{eqnarray}
In the case of the potential $W_0=\frac \lambda {x-x_0}$ it reduces to a
system of $N-1$ second order algebraic equations for $N$ variables - the
coefficients $a_i$ and parameter $\lambda $. As an example we now consider
the $N=4$ case. The coefficients $a_i$ are the solution of the system

\begin{eqnarray}
2a_1+2a_2a_3+4a_1a_2 &=&-\frac 1\lambda  \label{N4eq} \\
2a_2^2+2a_1^2+2a_2+2a_1a_3 &=&-\frac{a_1}\lambda  \nonumber \\
6a_1a_2+2a_3 &=&-\frac{a_2}\lambda  \nonumber
\end{eqnarray}
which can be turned into an equation of the 6th order and gives the roots
explicitly.Similarly, $N=3$ case gives three and $N=5$ fifteen roots. The
solutions could be relevant to the classification of the irreducible
representations of $S_N.$ It may turn out, that there exists the
correspondence between the $n$-particle conformal SUSY\ QM and $SU(n)$ YM in 
$D=2$ , which was discussed in \cite{Gorski}. Symmetry and classification of
the solutions for arbitrary $N$ requires an additional analysis. However,
there is one solution that can be found for any $N$. When $x_0$ is zero, it
corresponds to the known $N=1$ and $2$ conformal SUSY\ QM of \cite{Fubini}, 
\cite{Akulov} and its generalization for arbitrary $N$. In order to see that
we rewrite the potential as a complete square 
\begin{equation}
W^2+\sigma _1W^{\prime }=\frac{(\lambda P(\sigma _2,\sigma _3,...,\sigma _N)-%
\frac{\sigma _1}2)^2-(\frac{\sigma _1}2)^2}{x^2}  \label{Potsquare}
\end{equation}
The compatibility condition in this case can be written as 
\[
\lambda P(\sigma _2,\sigma _3,...,\sigma _N)-\frac{\sigma _1}2=\pm {\LARGE (}%
\lambda P(\sigma _1,\sigma _3,...,\sigma _N)-\frac{\sigma _2}2{\LARGE )} 
\]
For any $N$ it has a solution, which corresponds to the $+$ sign in the
equation above 
\begin{eqnarray}
a_i &=&0\qquad i\neq 1  \label{solut} \\
a_1 &=&-\frac 1{2\lambda }  \nonumber
\end{eqnarray}
Thus, the complete Hamiltonian of the conformal SUSY\ QM for any $N$ has
identical structure with that of $N=1$ (\ref{N1Ham} ), except for the
constant 
\begin{equation}
H_N=\frac 12\left( -\partial ^2+\frac 14\frac{\left( 2\lambda -\sigma
_N-...-\sigma _2\right) \left( 2\lambda -\sigma _N-...-2\sigma _1\right) }{%
x^2}\right)  \label{hamilt}
\end{equation}
The spectrum is obtained by redefining the constant $g$ in the spectrum of
the $N=0$ theory (\ref{Spectr}) with the result 
\begin{equation}
r_0=\frac 12(1+\lambda P(s_2,s_3,...,s_N)-\frac{s_1}2)\qquad r=r_0+n\text{%
\qquad }n=0,1,...
\end{equation}

\section{Discussion}

We have shown, that the extended SUSY QM can be constructed from the
assumption, that the Hamiltonian is the sum of the irreducible
representation of the symmetric group $S_N$. In order to satisfy all the
commutation and anticommutation relations in the simplest case (without
central charges) of the extended superalgebra, the parametric functions of
the representations must obey the compatibility condition. It has been shown
that these condition can be solved in the iteration procedure for an
arbitrary number of the supersymmetries $N$. The solution allows to
parametrize the Hamiltonian by a single arbitrary function. Using the
factorization property of the conformal potential, the conformal SUSY QM has
been derived for any $N$ as a natural generalization of the known $N=1$ and $%
N=2$ cases. However, it has been also found that there is the whole class of
possible generalizations of the theory for every $N$ corresponding to
various irreducible representations of the group $S_{N.}$

We would like to point out that the formalism can be broadened to include a
number of other aspects of the SUSY QM. Among a few, which seem to us most
interesting, we are going to discuss in the following. The incorporation of
the many dimensions \cite{Berezovoi} in this formalism will require the more
general extended superalgebra with the central charges. The equivalence of
the many dimensional and many particle theories in QM, as well as the
special role of the superconformal potential would make possible to
construct the SUSY QM with the Calogero type potentials for any $\mathcal{N}$%
. Such theory in $d=1$ and for $\mathcal{N}=4$ has been argued in \cite
{Gibbons} to be a microscopic description of a black hole. The question of
the superfield generalization of the extended SUSY QM has been left open.
There are exciting areas of the SUSY QM such as the SWKB and shape invariant
potentials, which can be treated using this formalism for arbitrary $N$. We
can anticipate the new results in the classification of the exactly solvable
potentials. In the context of the $N$ arbitrary we can consider a potential
of the SUSY QM as the expansion in the powers of the Plank constant $h$ and
introduce a new limit of the number of the supersymmetries $N$ approaching
infinity.

We would like to thank S. Catto for useful discussions and support. We are
also grateful R. Khuri , V.P. Nair, A. I. Pashnev, V. I. Popov and B.
Nicolescu for useful comments and discussions. This research has been
partially supported by PSC-CUNY Research Award. One of us (VA) has been
supported by grants INTAS 93-127ext, 93-493ext, 96-308.

\end{document}